\documentclass[aps,prl,twocolumn,superscriptaddress,amsmath,amssymb,nofootinbib]{revtex4-2}
\usepackage[utf8]{inputenc}
\usepackage{graphicx}
\usepackage{dcolumn}
\usepackage{bm}
\usepackage{epsfig}
\usepackage{float}
\usepackage{booktabs}
\usepackage{amsmath}

\usepackage{multirow}

\usepackage{mathtools}
\newtagform{normalsize}[\normalsize]{\normalsize(}{\normalsize)}
 
\begin{document}

\usetagform{normalsize}

\title{A first extraction of the weak magnetism form factor and Fierz interference term from the $^{114}$In $\rightarrow$ $^{114}$Sn Gamow-Teller transition} 

\author{L.~De~Keukeleere}
\email[Corresponding author: ]{lennert.dekeukeleere@kuleuven.be}
\affiliation{KU Leuven, Instituut voor Kern- en Stralingsfysica, Celestijnenlaan 200D, B-3001 Leuven, Belgium}
\author{D.~Rozpedzik}
\email[Corresponding author: ]{dagmara.rozpedzik@uj.edu.pl}
\affiliation{M. Smoluchowski Institute of Physics, Jagiellonian University, PL-30059 Cracow, Poland}
\author{N.~Severijns}
\affiliation{KU Leuven, Instituut voor Kern- en Stralingsfysica, Celestijnenlaan 200D, B-3001 Leuven, Belgium}
\author{K.~Bodek}
\affiliation{M. Smoluchowski Institute of Physics, Jagiellonian University, PL-30059 Cracow, Poland}
\author{L.~Hayen}
\affiliation{Normandie Univ, ENSICAEN, UNICAEN, CNRS/IN2P3, LPC Caen, F-14000, Caen, France}
\author{K.~Lojek}
\affiliation{M. Smoluchowski Institute of Physics, Jagiellonian University, PL-30059 Cracow, Poland}
\author{M.~Perkowski}
\affiliation{M. Smoluchowski Institute of Physics, Jagiellonian University, PL-30059 Cracow, Poland}
\affiliation{KU Leuven, Instituut voor Kern- en Stralingsfysica, Celestijnenlaan 200D, B-3001 Leuven, Belgium}
\author{S.~Vanlangendonck}
\affiliation{KU Leuven, Instituut voor Kern- en Stralingsfysica, Celestijnenlaan 200D, B-3001 Leuven, Belgium}

\begin{abstract}
Spectrum shape measurements in nuclear $\beta$ decay can be used to test physics beyond the Standard Model with results being complementary to high-energy collider experiments. In particular, Beyond Standard Model sensitivity of the weak interaction is expressed through the so-called Fierz interference term. Additionally, the $\beta$ spectrum shape is a useful tool to probe Standard Model effects, among which the most prominent is \textit{weak magnetism}, a higher-order recoil correction induced by nuclear pion exchange. To study effects in the $\beta$ spectrum shape at a precision level competitive with the LHC, a new spectrometer was designed and built. It consists of a 3D low-pressure gas electron tracker and a plastic scintillator used for triggering the data acquisition and recording the $\beta$ particle energy. 
In this Letter, the results from $\beta$ spectrum shape measurements on the allowed Gamow-Teller transition $^{114}\text{In} \rightarrow ^{114}\text{Sn}$ are presented, including a first extraction of the weak magnetism form factor in the high nuclear mass range and a new estimate of the $90\%$ confidence interval for the Fierz interference term.
\end{abstract}

\maketitle

The search for phenomena beyond the Standard Model (SM) is carried out on two energy frontiers. The high-energy frontier is represented by complex experiments performed at particle accelerators. At the low-energy frontier, new physics beyond the Standard Model is explored by high-precision experiments in nuclear and neutron $\beta$ decay, in particular, correlation coefficient measurements  \cite{Wauters2010,Markisch2019,Dubbers2021,3hasan,burkey2022,fenker2018} and recently also beta spectrum shape measurements \cite{Severin2014,Hughes2019,Birge2019,loidl2014}.~The aim of this work was to study both the Standard Model and New Physics beyond the Standard Model (BSM) through a precise spectrum shape measurement of the ground state-to-ground state $\beta^-$ pure Gamow-Teller transition $^{114}$In$(1^+)\rightarrow ^{114}$Sn$(0^+)$ with an endpoint energy $E_0 = 1989.93(30)$ keV \cite{Wang2021}. For this purpose, a small $\beta$ spectrometer called \textit{miniBETA} was built in an effort to avoid typical systematic effects limiting past precision $\beta$ spectrum shape measurements, e.g. backscattering and background phenomena.

In a $\beta$ spectrum shape measurement, the presence of BSM physics is indicated by a non-zero Fierz interference term $b_F$, 
with the sensitivity to exotic scalar and tensor currents being maximal for endpoints energies around $1-2$~MeV \cite{Gonzalez-Alonso2016}. In order to be competitive with high-energy and other low-energy experiments, a precision of the order of $10^{-3}$ is required. At this level of precision, even small Standard Model effects become important. As the decaying nucleon is embedded in a nucleus, one group of effects is induced by the strong interaction. This group is usually dominated by the so-called weak magnetism (WM) term \cite{Calaprice1976}.

In addition to BSM exploration, another important motivation for this work was to determine the size of the weak magnetism form factor for an isotope in the region of mass $A \approx 100$, where at present no data exist~\cite{Severijns2023}. Determining weak magnetism in the mass region of fission products is also of interest to the reactor antineutrino anomaly \cite{reactor1,reactor2,reactor3}, which is now evaluated using a nuclear structure-independent approach to account for weak magnetism. Moreover, the poor knowledge of weak magnetism has already limited the attainable precision of several of the recent BSM experiments or constituted a major contribution to their systematic uncertainty \cite{Pitcairn2009, Wauters2009, Wauters2010}. This is important for tensor current searches, which typically use pure Gamow-Teller transitions in order to optimize sensitivity (weak magnetism is absent in pure Fermi transitions). Recently an update was made of existing experimental knowledge on weak magnetism, combined with shell-model based calculations to gain further insight in its nuclear structure and/or mass dependence \cite{Severijns2023}.

To extract the WM form factor $b/Ac$ and Fierz term $b_F$ from the $\beta$ spectrum shape, a precise theoretical description is indispensable. A recent review \cite{Hayen2018,Hayen2019} summarized all SM effects playing a significant role in nuclear $\beta$ decay, e.g. radiative corrections and atomic and molecular effects were untangled and factorized to $\mathcal{O}(10^{-4})$. This description is used to calculate the theoretical spectrum, ignoring BSM physics and unknown SM physics such as weak magnetism. In absence of the former, the SM weak magnetism-dependent spectrum shape function reads\footnote{Compared to the theoretical description developed in Ref.~\cite{Hayen2018}, the spectrum was generated with a sign inversion in the second-order radiative correction according to Ref.~\cite{czarnecki2004}. A publication on this topic will follow.}:
\begin{equation}
\label{eq:fit_theory_spectrum}
	S_{th}(W,b/Ac) = \frac{C(W,b/Ac)}{C(W,0)}S_{th}(W,b/Ac=0),
\end{equation}
where $W$ is the electron's total energy in units of electron rest mass, $b$ is the weak magnetism form factor, $A$ is the mass number, $c$ the Gamow-Teller form factor, and the notation of Ref.~\cite{Calaprice1976} for the form factors was used. In light of BSM studies, it is sufficient to evaluate the ratio $b/Ac$, since weak magnetism only enters the spectrum shape description in this combination. Next, the nuclear shape factor $C(W)$ can be written for pure Gamow-Teller (GT) decay as a sum of terms having different energy dependencies~\cite{Hayen2018}:
\begin{equation}
\label{eq:shape_factor_rewritten}
\begin{split}
		C(W, b/Ac) \simeq  1+C_{0}(b/Ac)+C_{1}(b/Ac) W \\ + C_{-1}(b/Ac) / W +C_{2} W^{2} + \Phi \mathcal{P}(W).
\end{split}
\end{equation}
Here, $\Phi\mathcal{P}(W)$ denotes the pseudo-scalar contribution, which was found to be negligible in this case. Working out the remaining terms in Eq.~\ref{eq:shape_factor_rewritten} for the $^{114}$In$\rightarrow ^{114}$Sn transition, results in the following expression for the pre-factor in Eq. \ref{eq:fit_theory_spectrum}:

\footnotesize
\begin{equation}\label{eq:wm_spectral_effect}
    \frac{C(W,b/Ac)}{C(W,0)} = 1 + \frac{b}{Ac} \left(0.009 + 0.0007W - \frac{0.0004}{W}\right).
\end{equation}
\normalsize
\vspace{1mm}

Thus, weak magnetism acts on the $\beta$ spectrum in three ways. It changes the overall decay rate through the constant term and it changes the spectrum with both a linear and an inverse energy dependency. The two energy-dependent terms determine the shape of the spectrum, where the effect of the $1/W$ term is only significant in the low energy part of the spectrum. As a result, the relative effect of $b/Ac$ on the $\beta$ spectrum is primarily a slope with energy. The magnitude of this is approximately $\frac{0.0014}{\text{MeV}}$, i.e. over a range of 1 MeV one unit of $b/Ac$ will change the spectrum by $0.14\%$. Hence, it is at this level that systematics need to be kept under control. 

To compare the theoretical spectrum with the experimental one, the expression in Eq.~\ref{eq:fit_theory_spectrum} needs to be convoluted with a detector response function:

\footnotesize
\begin{equation}\label{eq:fit_detected_spectrum}
    \begin{aligned}
    S_{det}(ADC,b/Ac) &= \int_0^{Q} R(W,ADC)S_{th}(W,b/Ac)dW,
    \end{aligned}
\end{equation}
\normalsize

where $S_{det}(ADC,b/Ac)$ is the theoretical spectrum in ADC units, and $R(W,ADC)$ is the detector response, which for any given $W$ returns the posterior probability $P_R(ADC|W)$ of obtaining a particular ADC signal. The combination of a MWDC with a scintillator necessitates the inclusion of numerous physical processes in the response function $R(W, ADC)$ through Monte-Carlo simulations \cite{lennertPhD, minibetaPRC}. This involves a sampling of the real posterior probability distributions $P_R(ADC|W)$ for all energies between $0$ and $2$ MeV. Inevitably, this demands for a discretization of the energy range and a high-statistics sampling to obtain a close match with the underlying spectral response.

\begin{figure}[t]
    \centering
    \includegraphics[draft=false,width=64mm,height=62mm]{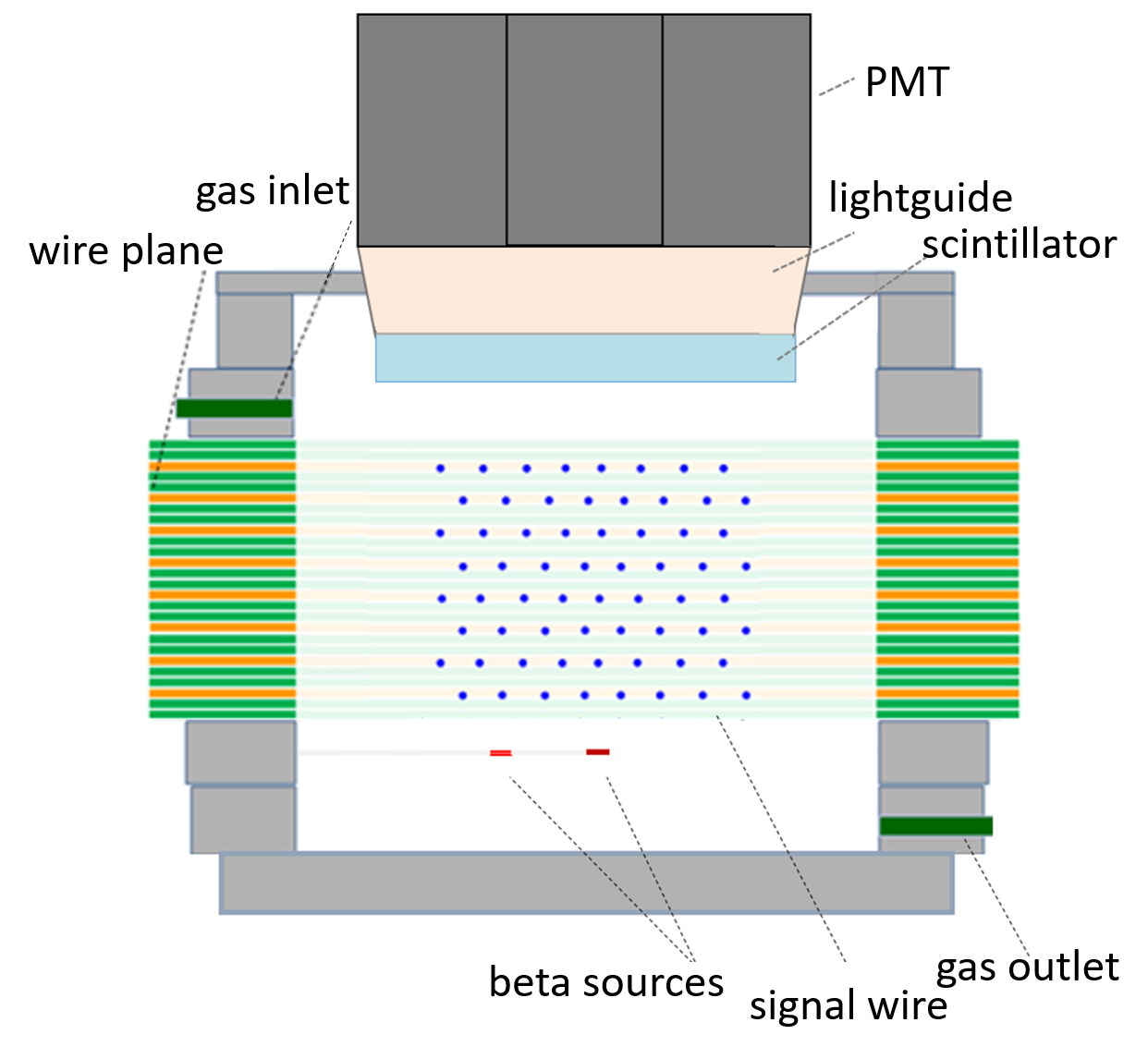} 
    \caption{The miniBETA spectrometer design.}
    \label{setup}
\end{figure}

The spectrum shape measurements were performed using
the \textit{miniBETA} spectrometer~\cite{miniBETADdaq, Maciejthesis}, which is a combination of a $\beta$ particle energy detector and a gas electron tracker optimized for low energy electrons. The sketch of the setup is shown in Fig.~\ref{setup}.
The former comprises a plastic scintillator disk with a diameter of $20$~cm and a thickness of $3$~cm, optically coupled to a light guide disk on which $4$ PMT's are stacked. The latter is a
multi-wire drift chamber (MWDC) having honeycomb-like cells with anode wires in the center at high voltages around $2000$ V, and grounded cathode wires at the corners. Based on earlier performance studies, i.e. comparing tracking resolutions and efficiencies, it was decided to use for the gas medium a mixture
of $70 \%$ helium and $30 \%$ isobutane at a pressure of $600$~mbar. The MWDC allows identifying events that distort the spectrum shape, e.g. electrons backscattering from the scintillator surface or cosmic muons traversing the experimental setup. In addition to event pattern recognition, the setup allows for several filtering and calibration procedures. For example, by requiring coincidence between the scintillator and drift chamber, external noise sources and gamma rays are filtered out. Furthermore, in order to correct for non-uniform light propagation in scintillator and light guide, tracking conversion electrons from a $^{207}$Bi calibration source, located next to the $^{114}$In source, enables the real-time generation of a 2D-detector surface gain map. Such a map is also useful for monitoring drift parameter effects unavoidable in long measurements.

A dedicated 3D-track reconstruction algorithm is used for recognizing event topologies and identifying different event origins. Electron tracking in the plane perpendicular to the wire is based on the standard drift time principle, while the track component along the wire is achieved by charge division method. As demonstrated in Fig. 2 of Refs. \cite{lennert2021,rozpedzik2022}, the applied data analysis methods are able to distinguish cosmic muons, $^{207}$Bi calibration electrons, $^{114}$In $\beta$ particles and backscattered electrons. The recognition efficiency was tested with MC simulations and found to be extremely good, giving rise to $>99$\% clean $^{207}$Bi electron conversion and $^{114}$In $\beta$ spectra. In addition, the algorithm is able to reject backscattered events off the scintillator with an efficiency of~$65$\%. As a proof of principle, the total experimental and simulated $^{207}$Bi and $^{114}$In $\beta$ spectra, with $b_{\text{F}}$ and $b/Ac$ put to zero, were compared.  The experimental spectra were found to be reproduced by simulation at the $10^{-2}$-level.
The $\beta$ spectrum comparison exhibits a $10^{-2}$-level slope in the fiducial energy region, i.e. $700-1800$ keV, then attributed to a possible non-zero $b/Ac$ ~\cite{lennert2021,rozpedzik2022}.

Aided by empirical data from calibration procedures, i.e. the MWDC cell-level calibration parameters and the position-dependent scintillator gain parameters, the detector response to electron energies ranging from 300 to 2000 keV was simulated. The weak magnetism form factor was then fitted by sampling the theoretical $\beta$ spectrum shape for varying $b/Ac$ values. To reduce the systematics involved with the energy response of the spectrometer, i.e. corrections in gain and quadratic effects, a bismuth-aided auto-calibration fit procedure was employed \cite{lennertPhD,minibetaPRC}, which was heavily inspired by the methods described in \cite{Huyan2018}.
The optimization procedure must now attempt to minimize the difference between $S_{det}$ and the real experimental spectrum $S_{exp}$. 
To account for a mismatch in normalization between these two spectra, one additional free parameter $r_N$, which corresponds to the ratio of the total counts, was added to the fit procedure. 

In a first stage, the self-consistency of the fitting routine was tested by feeding two theoretical $\beta$ spectra to the MC simulation: for the first one, $b/Ac$ was set to zero, while for the latter a first estimate of $8$ was plugged in. In addition, all known SM correction terms, as listed in Tab.~VII of Ref.~\cite{Hayen2018}, were included. The values extracted for $b/Ac$ were within the $1\sigma$ statistical uncertainty from the theoretical ones, proving the self-consistency of the extraction method.

\begin{figure}[t]
    \centering
    \includegraphics[draft=false,width=80mm,height=60mm]{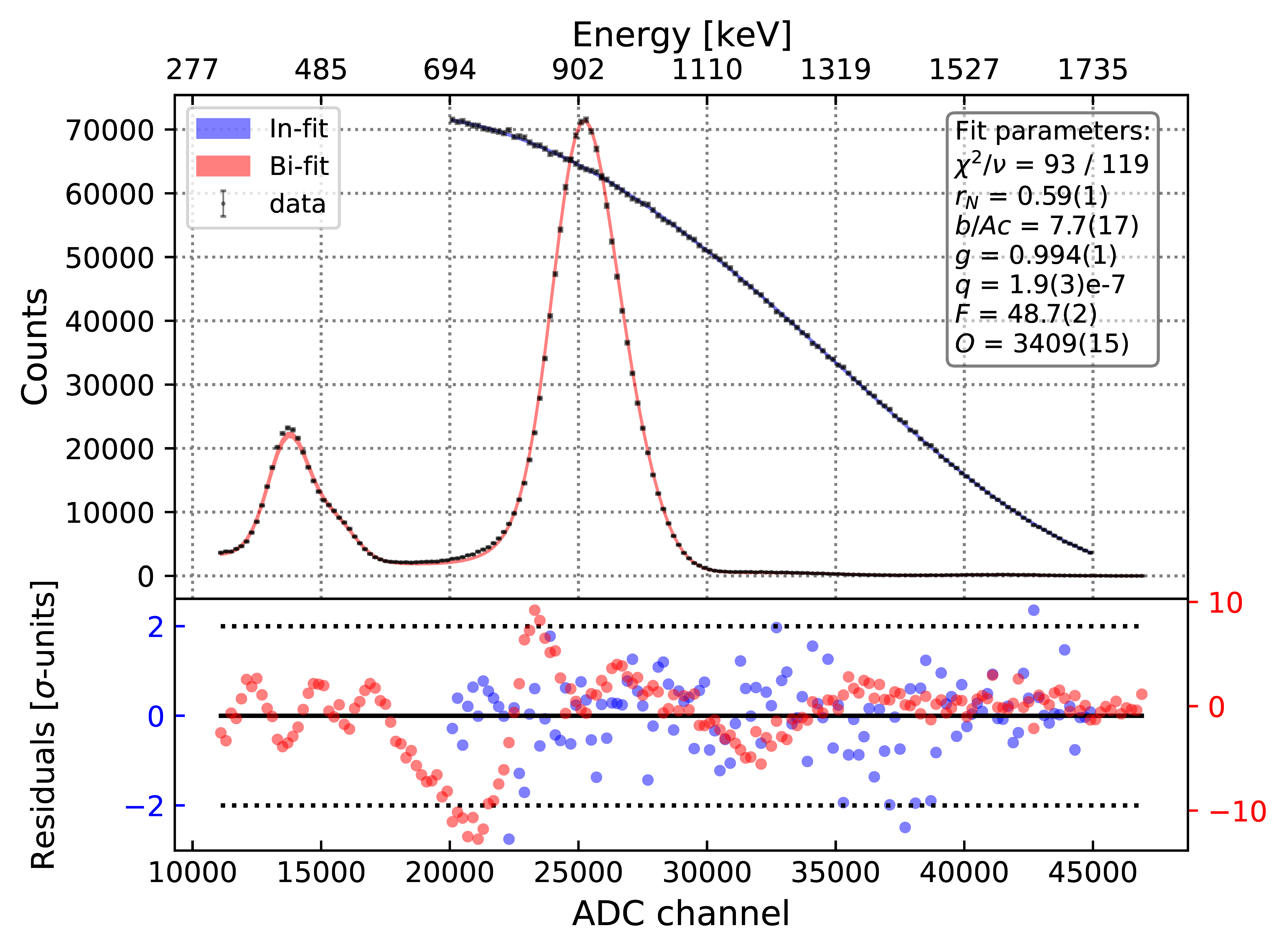} 
    \caption{The bismuth-aided auto-calibration fit result for the experimental data set. $g$ corresponds to the average gain correction, $q$ describes any quadratic effects in the conversion from energy to ADC, and $O$ and $F$ are the auto-calibration offset and resolution parameters. 
    Note that no systematic trend is seen in the residuals for the fit of the $^{114}$In data (blue points).}
\label{fig:hssx_60_hssz_60_minadc_20000_maxadc_45000_lmfit_recalib_no_fierz_no_systematics}
\end{figure}

Employing the fit algorithm to the experimental data of the first of four runs, leads to the determination of the value $b/Ac$= $7.7$, as presented in Fig. \ref{fig:hssx_60_hssz_60_minadc_20000_maxadc_45000_lmfit_recalib_no_fierz_no_systematics}. Analyzing also the data from three additional runs, gives a combined result of $b/Ac$= $7.1$. In a next step, the total uncertainty was estimated. 
The different contributions to the total uncertainty are listed and discussed below. More details will be given in a forthcoming paper (see also \cite{lennertPhD}).

\textit{Theory}: Two sub-leading form factors, i.e. the induced tensor form factor, $d$, and $\Lambda$, a matrix element ratio, were dropped form the nuclear shape factor terms ($C_{0,1,-1,2}$) in Eq. \ref{eq:shape_factor_rewritten} with respect to the original equations (\cite{Hayen2018}, Eq. 106). The form factor $d$ has shown to be either zero in case of $\beta$ transitions between analog states, or, in other cases, to be not bigger than the weak magnetism form factor $b$ \cite{Severijns2023}. Simulations show that, when using equal magnitudes for $b$ and $d$, i.e. a conservative approach, a 0.03 error on the fit result is introduced. In case of $\Lambda$, an extreme single-particle calculation (see Eq. 127 in \cite{Hayen2018}) yields a value of $1$. These estimates for the induced tensor form factor $d/Ac$, and $\Lambda$, lead to respective shifts in $b/Ac$ of $0.02$ and $0.23$. \newline
The pseudoscalar term is incorporated in the fit. Here, the choice was made to take a value for $\Phi$ midway between the PCAC free nucleon estimate and the estimate from a maximal quenching of $80\%$ on $g_P$. As expected from the relative magnitude of the linear energy term in $\mathcal{P}(W)$, the resulting $b/Ac$ shift is well below $0.1$. An additional error of $0.72$ is reserved for the uncertainty on the end-point energy ($0.3$ keV). Adding all in quadrature, gives a total error of $0.76$. 

\textit{Statistical/Fit}: The statistical uncertainty is the cumulation of three sources: (1) the statistics of the experimental $\beta$ spectrum, (2) the statistics of the simulated response data, and (3) the quality of the fit, including correlations between the fit parameters, predominantly between $r_N$ and $b/Ac$. As displayed in Fig. \ref{fig:hssx_60_hssz_60_minadc_20000_maxadc_45000_lmfit_recalib_no_fierz_no_systematics}, a fit uncertainty of $1.7$ is obtained for run 1. Combining all four runs (which were found to be
consistent within the $1\sigma$ error bars), a reduced uncertainty of $0.9$ was obtained.

\textit{Model}: The largest contribution to the systematic error budget originates from the bismuth-aided auto-calibration fit model, i.e. the assumption that the scintillator response as seen from the $\beta$ source is identical to the $^{207}$Bi calibration source. Foremost, the scintillator illumination by the two sources is slightly different, as they are in different locations. MC simulations were employed to evaluate the size of this effect. A shift of only $1$ unit in $b/Ac$ was observed. However, with the current level of simulation statistics, this result itself was subject to uncertainty, yielding an estimate of $\pm2$.

\textit{Material tolerances}: When including the experimental setup in the simulations that were used to create the detector response, ideal material properties were assumed. 
A few of these properties may distort the $\beta$ spectrum shape due to energy loss in the source foil, the MWDC gas mixture and scintillator reflector foil. MC simulations were employed, varying these material parameters, to obtain an uncertainty estimate of $\pm2$ on $b/Ac$. As this estimate is inherently subject to statistics, more simulations are underway to reduce this value.

\textit{Tracking}: The 3D track reconstruction method entails systematics of its own. This was extensively studied by comparison of experiment and simulation. It was concluded that within the chosen energy window, the spectrum distortion stays below $0.2\%$. To evaluate the effect on the extraction of $b/Ac$, the simulated spectral error bars were blown up by $0.2\%$, resulting in an uncertainty of $0.8$ induced by tracking.

\textit{Detector resolution}: The precision of the calibrated energy resolution was heavily influenced by low-energy electron tracking systematics. The calibration routine included two terms: a constant offset and an energy dependent term related to Poisson statistics. The quality of the fit, i.e. the distribution of residuals, was almost identical for different sets of these parameters. Nevertheless, the observed shift on $b/Ac$ amounted to only $0.5$.

Including all uncertainties listed above, the combined result 
for this first determination of the weak magnetism form factor 
now reads:
\begin{equation*}
    \begin{aligned}
        \frac{b}{Ac}&= 7.1\,(7)_{\text{th}}\,(9)_{\text{stat}}\,(20)_{\text{mod}}\,(20)_{\text{tol}}\,(8)_{\text{track}}\,(5)_{\text{res}}\\
        &=7.1\,\pm 3.2.
    \end{aligned}
\end{equation*}

Since an evaluation of $b/Ac$ in this mass range is an absolute first, a conservative approach was used to estimate the systematic uncertainties. Furthermore, there are currently no accurate theoretical predictions available for this uncharted territory. Nevertheless, it is interesting to note that the value of $7.1$ is within $1\sigma$ of the average experimental $b/Ac$ value determined in the $A<75$ mass range, i.e.  $5.1$ $\pm$ 2.4 (83 transitions)~\cite{Severijns2023}. Furthermore, assuming an extreme single particle configuration, a harmonic oscillator shape for the nuclear potential, and assuming the impulse approximation is appropriate for this transition, no orbital enhancement is anticipated. An effort from the theory side to calculate these matrix elements in a high mass nucleus, like $^{114}$In, could be worth the while, for a match would be a benchmark for both theory and experiment. Last but not least, it should be noted that the result presented above assumes absence of BSM physics.

To search for BSM physics, the procedure described above was extended to include the Fierz interference term $b_F$, where the BSM shape factor can be written as
\begin{equation}\label{eq:bsm_shape_factor}
	C_{\text{BSM}}(W) = 1 + \frac{b_F}{W}.
\end{equation}
The weak magnetism form factor $b/Ac$ for $^{114}$In exhibits a similar
energy dependency in the workable fit range. Nevertheless, a correlation study was performed on MC simulated data to investigate the BSM sensitivity~\cite{minibetaPRC}, indicating that a reasonable extraction was possible. 

\begin{table*}[t]
	\centering
	\caption{Overview of both published and unpublished Fierz estimates from $\beta$ asymmetry $A$ and $\beta$ spectrum shape ($SS$) experiments, with the first/second error giving the statistical/systematic uncertainty.}
	\resizebox{\textwidth}{!}{
    \footnotesize
		\begin{tabular}{cc|ll|c}
			\hline
			Source & Observable  & Fierz estimate (stat)(sys) unc.\hspace{5mm} & Total unc. & Project \\
			\hline \hline
			$n$ & $A$ & \hspace{10mm} $0.017(20)(3)$ & $0.020$ & PERKEO III \cite{Saul2020}\\
			$n$ & $A$ &  \hspace{10mm} $0.066(41)(24)$ & $0.048$ & UCNA \cite{Sun2020}\\
			$^{37}$K & $A$ &  \hspace{10mm} $0.033(84)(39)$ & $0.093$ & TRINAT \cite{Anholm2022}\\
			$^{20}$F & $S.S.$ &  \hspace{10mm} $0.0021(51)(84)$ & $0.0098$ & Michigan He$6$ group \cite{Hughes2019}(PhD, unpubl.) \\
			$^{45}$Ca & $S.S.$ &  \hspace{10mm} $0.40(5)(78)$ & $0.78$ & Los Alamos \cite{Birge2019}(PhD, unpubl.)\\
            $^{114}$In & $S.S.$ &  \hspace{10mm} $0.068(62)(29)$ &$0.068$ & miniBETA (this work) \\
			\hline \hline
		\end{tabular}
   \normalsize
    }
	\label{tab:fierz_measurements}
\end{table*}

In a next step, the bismuth-aided auto-calibration fit, with the addition of a seventh free parameter, representing $b_F$, was run on the experimental data. As expected, the uncertainty on $b/Ac$ increased significantly while the fit result for $b_F$  took a value of $0.062(117)$, with the reduced $\chi$-squared remaining unchanged. Including the three additional runs, a combined result of $0.068 (62)$ was obtained. Since the systematic effects are identical to the previous analysis, no elaborate discussion is needed. The size of each uncertainty source is added to the result, yielding
\begin{equation*}
    \begin{aligned}
        b_F &= 0.068\,(4)_{\text{th}}\,(62)_{\text{stat}}\,(13)_{\text{mod}}\,(24)_{\text{tol}}\,(5)_{\text{track}}\,(7)_{\text{res}}\\
        &=0.068\,\pm 0.068.
 \end{aligned}
\end{equation*}

Hence, the corresponding $90\%$ confidence interval is $-0.04<b_F<0.18$.\\
This result is in agreement with other experimental searches (see Tab.~\ref{tab:fierz_measurements}) and it is consistent with a zero value, confirming the SM. Furthermore, the large uncertainty is primarily attributed to the statistical correlation between $b_F$ and $b/Ac$. Simulations show that if $b/Ac$ were to be obtained independently, a competitive precision of about $0.01$ -- $0.02$ can be obtained for $b_F$.~Good candidates are the $T=1/2$ mirror transitions and transitions within a higher isospin multiplet ($T=1, 3/2, 2$), where CVC can be invoked to determine $b/Ac$~\cite{Severijns2023}. Hence, employing the miniBETA spectrometer to measure the spectrum shape for such a transition, could lead to an improved limit on BSM tensor couplings. This would require the setup to be installed, with small modifications, at a radioactive ion beam facility.

Another, independent determination of the beta spectrum shape of $^{114}$In has, in parallel with this work, been performed in the WISArD setup at ISOLDE-CERN with a totally different experimental approach using two scintillators in a strong magnetic field \cite{Simon}. Data analysis of this experiment is ongoing. 

The main goal of this research was to study the Standard Model and New Physics beyond it through a precise spectrum shape measurement of the $^{114}$In$(1^+)\rightarrow$ $^{114}$Sn$(0^+)$ $\beta^{-}$ decay.~The measurements were performed with a plastic scintillator in combination with a multi-wire drift chamber, where the latter served as an effective background filter. A first extraction of the weak magnetism form factor $b/Ac$ in a high-mass nucleus was achieved. In addition, an attempt was made to obtain a competitive uncertainty interval for the BSM Fierz term, yielding the third-best published result worldwide. \\

\begin{acknowledgments}
The experimental data were taken at Jagiellonian University.~We gratefully acknowledge the technical support provided by the IPJU mechanical workshop. This work has been supported by the BOF-KU Leuven Project GOA/15/010, Project G.0248.12 and G.0812.18N of the FWO Research Foundation Flanders.
\end{acknowledgments}

\end{document}